\documentclass[sigconf]{acmart}

\usepackage{svg}
\AtBeginDocument{%
  }

\setcopyright{acmlicensed}
\copyrightyear{2024}
\acmYear{2024}
\acmDOI{}

\acmConference[GenAIRecP Workshop '24]{Knowledge Discovery and Data Mining, Workshop on Generative AI for Recommender Systems and Personalization '24}{August 25--26,
  2024}{Barcelona, Spain}
\acmISBN{}

\begin{document}

\title{EDGE-Rec: Efficient and Data-Guided Edge Diffusion For Recommender Systems Graphs}


\author{Utkarsh Priyam}
\authornote{All authors contributed equally to this research.}
\email{upriyam@cs.cmu.edu}
\author{Hemit Shah}
\authornotemark[1]
\email{hemits@cs.cmu.edu}
\author{Edoardo Botta}
\authornotemark[1]
\email{ebotta@cs.cmu.edu}
\affiliation{%
  \institution{Carnegie Mellon University}
  \city{Pittsburgh}
  \state{Pennsylvania}
  \country{USA}
}


\begin{abstract}
  Most recommender systems research focuses on binary historical user-item interaction encodings to predict future interactions. User features, item features, and interaction strengths remain largely under-utilized in this space or only indirectly utilized, despite proving largely effective in large-scale production recommendation systems. We propose a new attention mechanism, loosely based on the principles of collaborative filtering, called Row-Column Separable Attention (\textbf{RCSA}) to take advantage of real-valued interaction weights as well as user and item features directly. Building on this mechanism, we additionally propose a novel Graph Diffusion Transformer (\textbf{GDiT}) architecture which is trained to iteratively denoise the weighted interaction matrix of the user-item interaction graph directly. The weighted interaction matrix is built from the bipartite structure of the user-item interaction graph and corresponding edge weights derived from user-item rating interactions. Inspired by the recent progress in text-conditioned image generation, our method directly produces user-item rating predictions on the same scale as the original ratings by conditioning the denoising process on user and item features with a principled approach.
\end{abstract}

\ccsdesc[500]{Computing methodologies~Machine learning algorithms}
\ccsdesc[500]{Computing methodologies~Probabilistic reasoning}

\keywords{recommendation, diffusion, graph, recommender, systems}

\received{29 July 2024}

\maketitle

\section{Introduction}
Denoising diffusion models \citep{ho2020denoising} form a prominent class of models that have proven remarkably effective across a variety of generation and sampling tasks where the data distribution is high-dimensional, such as video, image synthesis and graph generation \citep{liu2023generative}. However, generative modeling of graphs has historically focused on smaller graphs for applications such as molecule generation \cite{vignac2023digress, liu2023generative}. Recent work involving graph generation with diffusion models show promising results for reconstructing large undirected graphs with thousands of nodes \cite{wang2023diffusion, bergmeister2024efficient}. Prior work also presents promising results with guided denoising steps incorporating score functions to evaluate graphs generated during denoising \cite{jo2022scorebased}.

Recommendation systems focus on using historical user-item interaction data (such as reviews) to predict/recommend new items for users. Previous research in this space applies diffusion models to augment sparse data, model user dynamics/preference distributions, model item latent distributions, or perform sequential recommendation \cite{wang2023diffusion, yang2023generate, ma2024plugin}. In contrast to prior research in diffusion-based recommendation systems, we aim to construct a recommender system based on graph diffusion models by formulating the problem as a graph completion task. We focus primarily on the graph structure of the user-item interaction space.

\begin{figure}[h]
    \centering
    \vspace{-1.5pc}
    \includegraphics[width=\linewidth]{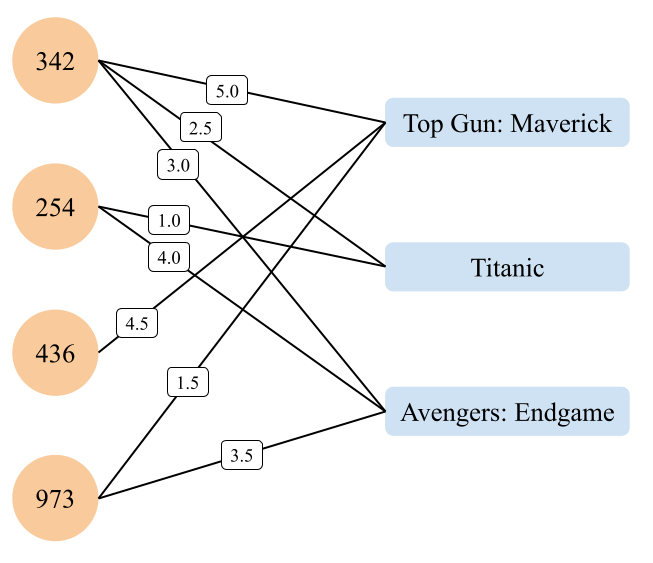}
    \caption{User-item interaction graph for a movie rating dataset. Each edge weight corresponds to the user-provided rating of the movie. Users and movies are linked via only their rating interactions, hence the graph is bipartite.}
    \Description{Bipartite graph of orange user nodes with numbered ID's, and blue movie nodes with names. Lines connect users to movies with one box interrupting each edge with the rating given by the user for the movie.}
    \label{fig:rating_graph}
\end{figure}

Our proposed method involves representing historical user-item interactions as edges between nodes in a bipartite graph with real-valued edge weights. These weights can represent the strength of these interactions (e.g. a rating out of 5). We also utilize latent space representations of users and items to condition the denoising process for the prediction of new interaction links (recommendations). In recommendation systems, the set of users and items (as well as their attributes) is fixed, so we focus on noising edge attributes of the interaction graph such as the adjacency matrix and edge weights in a continuous manner. Additionally, since recommender systems graphs are bipartite, we can represent the edge weights between users and items in a weighted interaction matrix, forming a 2D array similar to an image where each pixel represents the corresponding interaction strength between a user and item.

To the best of our knowledge, our work is the first to consider the interaction graph in a recommender system with both node features (per user and item) and the strength of interactions as edge weights. By conditioning on node features in the denoising model, we are able to guide the diffusion of weights over all possible interaction edges in the graph to approximate the expected interaction strengths. 

\section{Related Work}

\subsection{Diffusion on Adjacency Matrices}
In the context of graph generation, a natural formulation of a diffusion model defines the noising process in the space of adjacency matrices. In this spirit, EDGE \citep{chen2023efficient} defines the transition kernel of the forward process as a Bernoulli distribution that resamples each edge with a time-varying positive probability. It aims to reconstruct the original graph by removing edges in the forward process, and denoising by adding edges in the reverse process. By explicitly modelling the node degrees and appropriately restricting each step of the backward process to a selected subset of ``active'' nodes, EDGE achieves promising performance and efficiency on larger graphs. 


\subsection{Diffusion Transformers}
Transformer based denoising models (those that predict the noise added in a forward step by $q(x_t|x_{t-1})$ or predict $x_0$ directly), have become the standard for image diffusion models. Considering our aim to perform continuous diffusion on a 2D weighted interaction matrix with similar characteristics to an image, our denoising model is inspired heavily by the architecture utilized in the recent work on Diffusion Transformers \cite{peebles2023scalable}. Our main goal is to capitalize on the modeling capacity of scaled dot-product attention to capture neighborhood characteristics for each node in the interaction graph. We follow a similar approach to the original Diffusion Transformer architecture with regards to the usage of multi-head attention and time-step conditioning via adaptive layer-norms.

\subsection{Diffusion-based Recommender Systems}

Due to the structure implied by user-item interactions, recommendation problems can be naturally reformulated as link prediction problems on graphs. Applications of graph convolutional networks \citep{wang2021graph} to this task have been extensively studied, while applications of diffusion models here are relatively understudied. Recently, \citet{wang2023diffusion} proposed DiffRec, a recommendation paradigm based on generative diffusion models. Given a user and a list of items, the diffusion process acts on the space of the interaction vectors. That is, it reconstructs one-hot encoded vectors representing whether the user has interacted with a specific item or not. 

\subsection{Graph Diffusion Recommender Systems}

Subsequently, \citet{jiang2023diffkg, zhu2023graphaware} have incorporated the graph structure of the interaction history, bridging the gap between recommendation systems and graph diffusion models. However, they still do not fully utilize user or item features, and perform the forward and reverse diffusion processes over the space of one-hot encodings for interactions, regardless of interaction strength.

A concurrent work by \citet{yi2024directional} formulates the problem of recommendation similarly, however, they first train a graph neural encoder to produce node (user/item) embeddings. In their experiment with continuous diffusion, they noise these embeddings in the forward process, and train a denoising model to recover the embeddings. Their graph neural encoder relies mainly on the graph defined by historical user-item interactions, and they additionally incorporate user/item features by measuring the similarity between user-user and item-item pairs based on their features. 

Our work differs from their approach as we incorporate user and item features directly in our denoising model architecture rather than producing node embeddings using this information. While they perform diffusion over the space of these node embeddings, our diffusion process is applied to the space of weighted interaction matrices. As a result, our denoising model is able to directly predict interactions strengths whereas theirs recover node embeddings.


\begin{figure}[h]
    \centering
    \includegraphics[width=\linewidth]{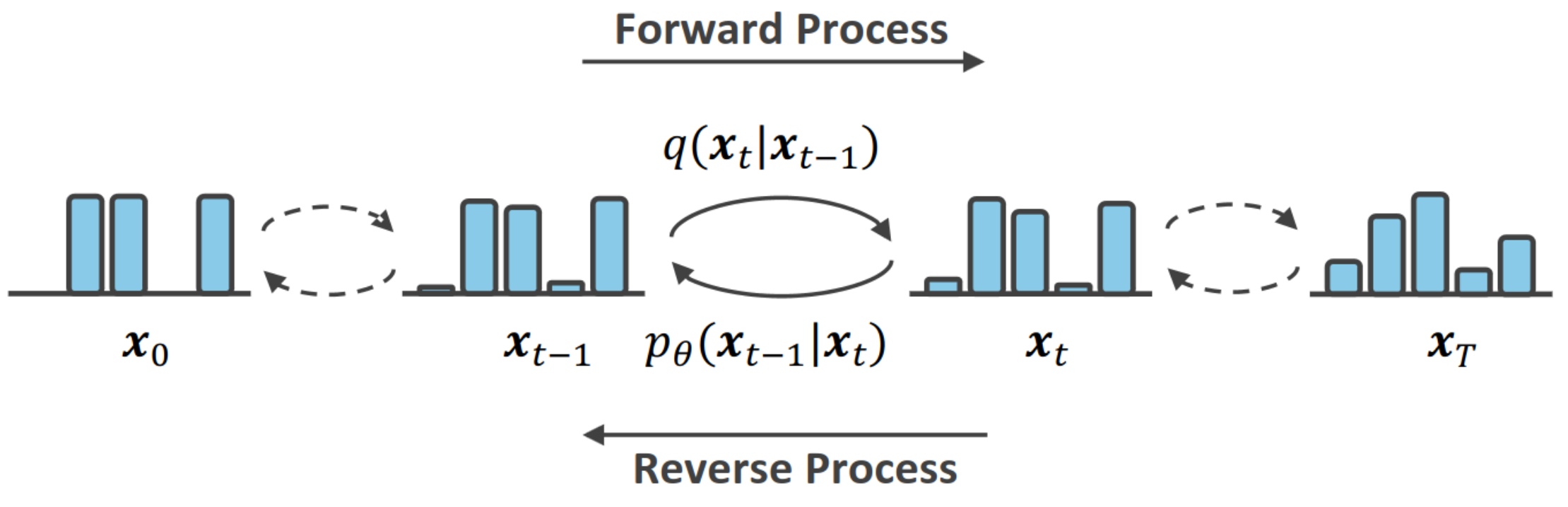}
    \caption{DiffRec - Starting from a one-hot vector of historical interactions, the forward process noises the user's interaction history until timestep $T$ by the transition step $q(x_t|x_{t-1})$. The model is trained to recover $x_0$ using $p_\theta(x_{t-1}|x_t)$ \cite{wang2023diffusion}.}
    \Description{Visual overview of diffusion process where a one-hot vector is visualized as a bar graph. Arrows in the forward direction point to new bar graphs representing weights in the one-hot vector being noised. Arrows in the reverse direction point to the previous denoised bar graph representing the one-hot encoded vector.}
    \label{fig:DiffRec}
\end{figure}

\section{Preliminaries}

In the traditional formulation of denoising diffusion models \cite{ho2020denoising}, the forward diffusion process gradually adds Gaussian noise to the original data, $\mathbf{x}_0 \sim q(\mathbf{x})$, over $T$ diffusion steps to reach random Gaussian noise, $\mathbf{x}_T \sim \mathcal{N}(0, \mathbf{I})$. Each diffusion step corrupts the noised representation from the previous step: $$q(\mathbf{x}_t | \mathbf{x}_{t-1}) = \mathcal{N}(\mathbf{x}_t; \sqrt{1-\beta_t} \mathbf{x}_{t-1}, \beta_t \mathbf{I})$$ Where $\{\beta_t\}_{t=1}^T$ are the noise-schedule hyper-parameters controlling the amount of noise added at each step. Note $\beta_t \in (0,1)$.

\citet{ho2020denoising} also provide the analytical form for $\mathbf{x}_t$ given $\mathbf{x}_0$:
\begin{align*}
    q(\mathbf{x}_t | \mathbf{x}_0) &= \mathcal{N}(\mathbf{x}_t; \sqrt{\widebar{\alpha}_t} \mathbf{x}_0, (1-\widebar{\alpha}_t)\mathbf{I}) \\
    &= \sqrt{\widebar{\alpha}_t} \mathbf{x}_0 + \sqrt{1-\widebar{\alpha}_t}\epsilon, & \epsilon\sim\mathcal{N}(0,\mathbf{I})
\end{align*}
Where $\alpha_t = 1-\beta_t$ and $\widebar{\alpha}_t = \Pi_{i=1}^t \alpha_i$. This enables any arbitrary step of the diffusion process to be generated extremely quickly for any given $\mathbf{x}_0 \sim q(\mathbf{x})$, simply by sampling $\epsilon\sim\mathcal{N}(0,\mathbf{I})$. The diffusion steps can be tuned by choosing an appropriate set of $\{\beta_t\}_{t=1}^T$.

The reverse diffusion process aims to gradually recover the original data $\mathbf{x}_0$ given the fully corrupted $\mathbf{x}_T$. \citet{ho2020denoising} show that the posterior Gaussian, $p(\mathbf{x}_{t-1} | \mathbf{x}_t, \mathbf{x}_0)$, is tractable (when conditioned on $\mathbf{x}_0$) and has the analytical form:
\begin{align*}
    p(\mathbf{x}_{t-1} | \mathbf{x}_t, \mathbf{x}_0) = &\mathcal{N}(\mathbf{x}_{t-1}; \mu_{t}(\mathbf{x}_t, \mathbf{x}_0), \tilde{\beta_t}\mathbf{I}) \\
    \mu_{t}(\mathbf{x}_t, \mathbf{x}_0) := &\frac{\sqrt{\widebar{\alpha}_{t-1}}\beta_t}{1-\widebar{\alpha}_t}\mathbf{x}_0 + \frac{\sqrt{\alpha_n}(1-\widebar{\alpha}_{t-1})}{1-\widebar{\alpha}_t}\mathbf{x}_t \\
    \tilde{\beta}_t := &\frac{1-\widebar{\alpha}_{t-1}}{1-\widebar{\alpha}_t}\beta_t
\end{align*}

Above, the conditioning on $\mathbf{x}_0$ is necessary for the posterior to be tractable. The primary goal of a diffusion model is to learn the initial distribution such that arbitrary Gaussian noise can be gradually denoised to produce outputs from the original input data distribution $q(\mathbf{x})$. Without conditioning on $x_0$, we need:
$$p_\theta(\mathbf{x}_{t-1} | \mathbf{x}_t) = \mathcal{N}\left(\mathbf{x}_{t-1}; \mu_\theta(\mathbf{x}_t, t), \Sigma_\theta(\mathbf{x}_t, t)\right)$$

The model, parameterized by $\theta$, is trained to approximate either the mean, $\mu_\theta$, and variance, $\Sigma_\theta$, \textit{or} to predict the added noise, $\epsilon$. Generally, the latter approach has lower variance and this parameterization more closely resembles Langevin dynamics with a loss that resembles the denoising score matching objective \cite{ho2020denoising}:
$$\mathcal{L} = \left\|\epsilon - \epsilon_\theta(\mathbf{x}_t, t)\right\|^2 = \left\|\epsilon - \epsilon_\theta\left(\sqrt{\widebar{\alpha}_t}\mathbf{x}_0 + \sqrt{1-\widebar{\alpha}_t}\epsilon, t\right)\right\|^2$$
where $\epsilon_\theta(\cdot)$ is the noise predicted by the model. We use this vanilla diffusion approach over the space of weighted interaction matrices (applying isotropic noise in the forward steps, and training our GDiT architecture to predict the added noise in the reverse process). Inspired by text-conditioned image-generation methods \cite{rombach2022high}, our approach also conditions on user features, $\mathcal{U}$, as well as item features, $\mathcal{I}$, such that the model predicts $\epsilon_\theta(\mathbf{x}_t, t, \mathcal{U}, \mathcal{I})$ and learns the posterior distribution $p_\theta(\mathbf{x}_{t-1} | \mathbf{x}_t, \mathcal{U}, \mathcal{I})$.

\subsection{Adjacency Matrix vs. Interaction Matrix}

With a bipartite undirected graph, there is no need to encode the edges with a full adjacency matrix. We instead define our diffusion process to operate over weighted interaction matrices consisting of users along the rows and items along the columns. For example, for the small interaction graph provided in \autoref{fig:rating_graph} we show below both the adjacency matrix and the weighted interaction matrix (abbreviations are used in place of the full movie titles).

\begin{figure}[H]
    \tiny
    \tt
    \centering
    \begin{tabular}{c|c|c|c|c|c|c|c|}
         & 342 & 254 & 436 & 974 & TGM & TTN & AVG \\ \hline
     342 &     &     &     &     & 5.0 & 2.5 & 3.0 \\ \hline
     254 &     &     &     &     & --- & 1.0 & 4.0 \\ \hline
     436 &     &     &     &     & --- & --- & 4.5 \\ \hline
     974 &     &     &     &     & 1.5 & --- & 3.5 \\ \hline
     TGM & 5.0 & --- & --- & 1.5 &     &     &     \\ \hline
     TTN & 2.5 & 1.0 & --- & --- &     &     &     \\ \hline
     AVG & 3.0 & 4.0 & 4.5 & 3.5 &     &     &     \\ \hline
    \end{tabular}
    \rm
    \caption{Full adjacency matrix for the interaction graph provided in \autoref{fig:rating_graph}. Notice the empty quadrants in the top left and bottom right of the matrix due to the bipartite graph.}
    \Description{Adjacency matrix filled with edge weights for bipartite graph of users and movies with edges corresponding to ratings between 1 to 5. Cells corresponding to edges between user-user nodes or movie-movie nodes are empty as they do not exist in the bipartite graph, whereas cells where the user has not rated the corresponding move are filled with three dashes.}
    \label{fig:adj_matrix}
\end{figure}

\begin{figure}[H]
    \tiny
    \tt
    \centering
    \begin{tabular}{c|c|c|c|}
         & TGM & TTN & AVG \\ \hline
     342 & 5.0 & 2.5 & 3.0 \\ \hline
     254 & --- & 1.0 & 4.0 \\ \hline
     436 & --- & --- & 4.5 \\ \hline
     974 & 1.5 & --- & 3.5 \\ \hline
    \end{tabular}
    \rm
    \caption{Weighted interaction matrix for the same graph (with non-existent interactions as `{\tt ---}'). Our proposed diffusion method operates over these weighted interaction matrices.}
    \Description{Weighted interaction matrix structure which is consistent with the top right corner of the adjacency matrix of the user-movie rating graph. Users are on the vertical axis and movies along the horizontal axis, with ratings in the cells of the array.}
    \label{fig:interaction_matrix}
\end{figure}

\subsection{Recommender System Constraints}
Formulating the recommendation task as a graph diffusion problem introduces two key constraints. First, the set of all users and items (as well as their metadata/attributes) is fixed (e.g. Netflix's subscribers and movie library). Second, the scale of the graphs we aim to operate on is on the order of at least $10^3$ nodes and $10^5$ edges. This excludes the majority of prior work on generating graphs using diffusion models as they operate on nodes, edges, and attributes in the forward and reverse processes, and they cannot scale to graphs with thousands of nodes (often because the methods are geared towards applications like molecule generation, where the graphs are substantially smaller and denser than recommender systems). 


\section{Proposed Framework}



We take inspiration from the notion of ``active node'' sampling in the interaction graph from EDGE \cite{wang2023diffusion}, thus enforcing the same mechanism of scalable local diffusion. Below, we highlight some key differences between our method and prior works.
\begin{enumerate}
    \item Sets of active nodes are sampled a priori and kept fixed throughout the diffusion. This means that our diffusion process is defined on the space of $n \times m$ sub-matrices (called \textbf{``patches''}), whose values represent all ratings in a sub-sample of $n$ users and $m$ items from the dataset. During training this translates to a batch of $n\times m$ interaction matrix patches which are corrupted in the forward process. The model is trained to predict the added noise within these patches given the corrupted data, time-step, and feature information for the $n$ users and $m$ items within each patch.
    \item For all user-item pairs within a sampled $n\times m$ patch, we consider real-valued \textbf{interaction strengths} as opposed to binary interactions (e.g. ratings from 1-5 for the MovieLens datasets \cite{movielens}). This makes the task more difficult but at the same time closer to applications of recommendation systems. The interaction strengths are mapped to the continuous interval $[-1,1]$, with $0$ representing a neutral interaction, and positive and negative values corresponding to favorable and unfavorable interactions, respectively.
    \item  With a large (and dense) enough dataset it is reasonable to assume that the distribution of these interaction strengths will be roughly Gaussian, or can easily be transformed to a standard Gaussian in an invertible manner, such as with a quantile transformation. This motivates the usage of the vanilla diffusion method over the space of (transformed) weighted interaction matrices. 
    \item In contrast to EDGE, we noise all edges within a patch at the same time, without employing additional selection at each step. This is made computationally feasible by the a priori subsampling of the active edges, which makes each subsample of size significantly smaller than the entire graph.
    \item  In contrast to DiffRec, we contract the weights of the interaction graph edges to be samples from a standard Gaussian distribution. While DiffRec stops the forward process at some partially-noised time-step to encode historical interaction information as a latent prior, we hope to remove this requirement by conditioning on user and item features, enabling generation by sampling directly from random noise.
\end{enumerate}


\subsection{Row-Column Separable Attention (RCSA)}

A key aspect of weighted interaction matrices is that the axes ordering of users and items can be permuted arbitrarily without changing the encoded information. As a result, spatial correlations in an interaction matrix are not analogous to those present in images. Due to this fact, the predictive power present within a square neighborhood of a user-item pair is limited. 

We hypothesize that most of the predictive power relevant to the noise prediction at a particular user-item pair is contained in the interaction matrix entries with either the same user or the same item, that is across the same row or the same column. Analogously, we hypothesize the amount of predictive power present across user-item pairs with different items and different users to be low. 

\begin{figure}
    \centering
    \includegraphics[width=0.8\linewidth]{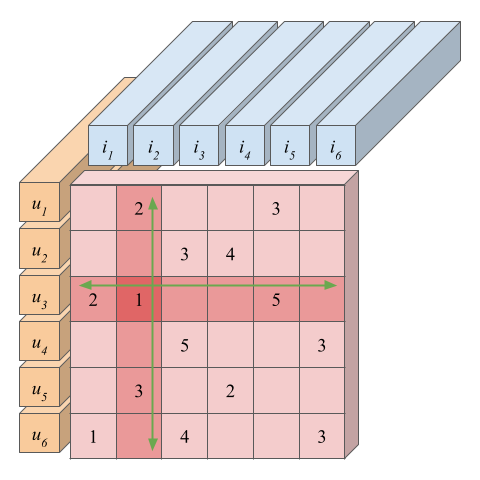}
    \caption{A submatrix/subsample from the full weighted interaction matrix including additional features provided to the denoising model (latent representations for user features, $u_i$, and item features, $i_j$). We also show our custom attention mechanism's operation for the user-item pair $(u_3, i_2)$, attending over other users' ratings for the same item (column), as well as the same user's ratings of other items (row).}
    \Description{A visualization of the submatrix from a weighted interaction matrix as a grid of red three dimensional cubes. Users are again along the vertical axis and items along the horizontal axis. Additionally, each user and item is depicted along their respective axes as a vector in three dimensions in the depth dimension to represent the features of each user and item. For row three and column two, these rows are highlighted in darker red with green arrows stretching across the row and column to represent the attention mechanism applied for the user three, item two pair.}
    \label{fig:rowColAttn}
\end{figure}

The presence of this structure suggests that the commonly used U-Net \cite{ronneberger2015u} architecture based on square convolutions is not suitable as it would only capture information within an immediate neighborhood. For this reason, we propose a novel attention module for interaction matrices. Within this module, each element of the matrix is separately attending to elements in its row first and elements in its columns subsequently. This is equivalent to a two-headed masked attention module where the attention coefficients for elements that do not belong to the same row or column, respectively, are masked out. In practice, one attention head operates on the sub-sample of the weighted interaction matrix and the other operates on its transpose. Replacing all attention modules with RCSA modules reduces the computation cost for an $N\times M$ patch from $O(N^2M^2)$ for traditional 2D attention mechanisms to $O(N^2+M^2)$.


\subsection{Graph Diffusion Transformer (GDiT)}

\begin{figure}
    \centering
    \includegraphics[width=0.87\linewidth]{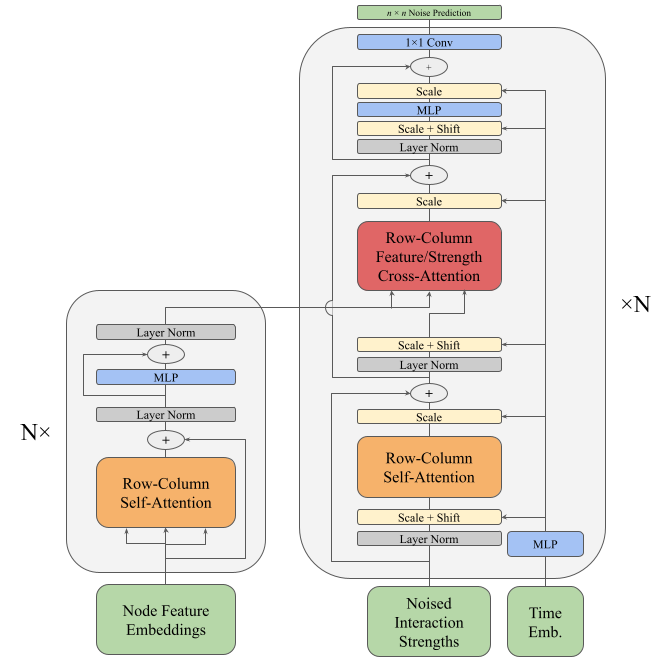}
    \caption{Novel GDiT Architecture}
    \label{fig:gdit_arch}
    \Description{Architecture of the novel graph diffusion transformer. From the bottom left a block of operations is defined which accepts node feature embeddings as inputs. These inputs pass through a row column self attention block, residual connection from the input, layer norm, multi-layer perceptron, residual connection from the output of the layer norm, and a final layer norm. The output of these blocks is connected to a row column cross-attention block. On the right a block of operations is defined to accept noised interaction strengths, and diffusion time step embeddings. The time embeddings are passed through a multi-layer perceptron to generate scale and shift parameters. The noised interaction strengths pass through a layer-norm, scale and shift, row column self attention block, scale, residual connection from the input, layer norm, scale and shift, row column feature strength cross attention, scale, residual connection from prior to the last layer norm, layer norm, scale and shift, multi-layer perceptron, scale, residual connection from prior to the last layer norm, 1x1 convolution, to produce the noise prediction outputs.}
\end{figure}

We propose a new transformer architecture (\autoref{fig:gdit_arch}) based on Diffusion Transformers (DiT) \cite{peebles2023scalable}, which we use to attend to user and item features in predicting diffusion noise. Specifically, we merge DiT with scaled layer norms conditioned on the diffusion time step and DiT with cross-attention conditioned on node features. We also apply RCSA (\autoref{fig:rowColAttn}) in place of all 2D attention operations. This architectural change better suits attention on bipartite graph structures, as well as providing runtime efficiency by replacing 2D attention with 1D attention. Note that each block can be repeated an arbitrary number of times to enhance the model's learning capacity for higher dimensional node feature embeddings (or be used to predict multi-dimensional rating/interaction strengths).

\section{Evaluation}
We train and evaluate our proposed method, EDGE-Rec, on the MovieLens datasets ML-100k and ML-1M \cite{movielens}. Evaluation is performed using a set of standard metrics for recommendation systems based on the top-$K$ predictions from the model. We train each model for at least 10000 iterations with a batch size of 16 and patch size at least $50\times 50$. We construct a custom time-aware 90-10 train-test split of the edges to ensure consequentiality and the absence of overlap in the time ranges covered by training and validation split.

Training consists of sampling random patches of specified size $n\times m$ from the target graph. Uniform random sampling of the full weighted interaction matrix over at least 10000 iterations allows most if not all of the users and items to be included during training. The model is provided patches corrupted to uniformly sampled time-steps, the time-steps each patch is noised to, and user-item features for each patch. Unknown interactions are optionally masked, and, if a user has not interacted with an item, the expected or true interaction strength is set to 0 (neutral).


Finally, we train the GDiT model by maximizing the usual denoising formulation of the ELBO objective, which we regularize at the single step level using Bayesian Personalized Ranking loss \cite{rendle2012bpr} to encourage coherent rankings. At sampling time, we reconstruct patches of ratings from random noise. Notably, given a patch we have access to the known training edges, which precede the test edge edges. This allows to frame the sampling problem as a \textbf{matrix completion} problem. Inspired from \textbf{inpainting techniques} for image completion \cite{lugmayr2022repaint}, we replace the known portion of the patch with the known noised ratings from the training set to further condition the generation on the a priori information and ecourage accurate generation of new ratings that ``complete'' the already-available ones. These candidates are discarded at test time.

We also propose a new method of denoising larger subgraphs by randomly tiling the encompassing subgraph with smaller patches at each reverse-diffusion step. As a result, the entire interaction matrix is denoised uniformly during the sampling process, instead of independently as when patches are naively denoised sequentially.

\begin{figure}
    \centering
    \includegraphics[width=1\linewidth]{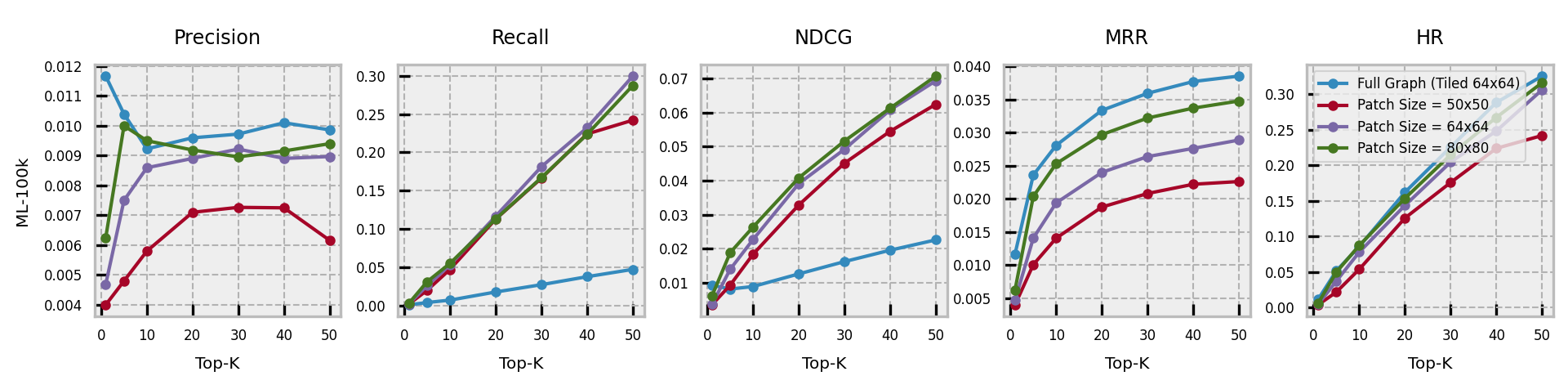}
    \caption{EDGE-Rec 8-block model evaluation metrics on ML-100k dataset. The figures show average results on arbitrary randomly sampled patches of varying sizes, as well as tiled sampling at $64\times64$ over the entire user-item rating graph.}
    \Description{Comparison of the evaluation metrics for an 8-block model following the previously described architecture plotted as a set of line graphs. Along the horizontal axis, multiple plots depict the trend in precision, recall, normalized discounted cumulative gain, mean reciprocal rank, and hit rate as K is increased to take the model's top K predictions for the ML-1M dataset. There are four lines per plot, showing the trends in results for each metric as K is increased, for 3 sampling patch sizes $50\times50$, $64\times64$, and $80\times80$, as well as tiled sampling over the full graph with tile size $64\times64$.}
    \label{fig:edgerec_random_subgraph_metrics}
\end{figure}

\begin{figure}
    \centering
    \includegraphics[width=1\linewidth]{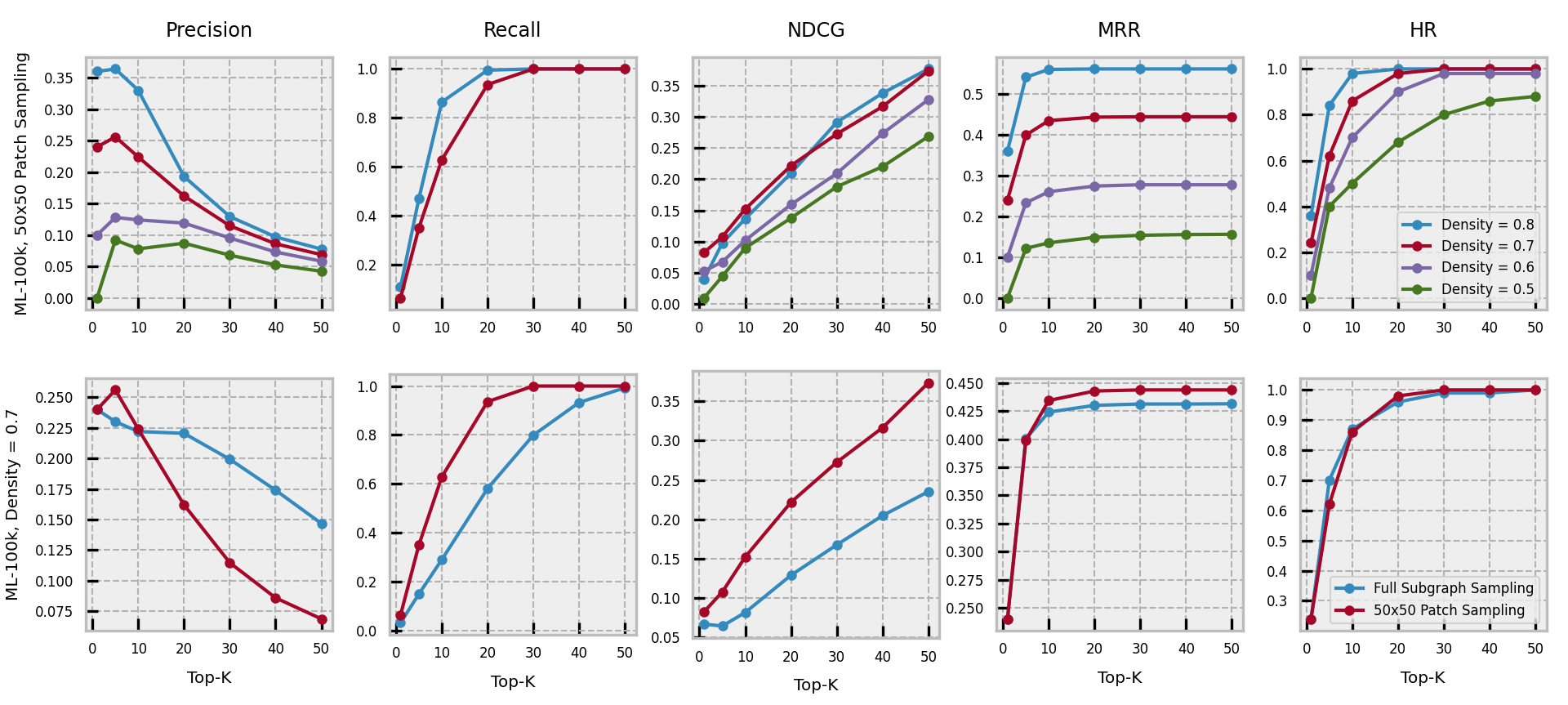}
    \caption{EDGE-Rec 1-block model evaluation metrics on ML-100k dataset. The top row depicts results on arbitrary $50\times50$ patches from subgraphs of specified density. The bottom row compares evaluations for sampling a fixed $50\times50$ patch to sampling the entire subgraph for a fixed label density of $70\%$, which corresponds to the top 100 users and 100 movies, by rating density (i.e. a $100\times100$ patch).}
    \Description{Evaluation metrics for the previously described architecture plotted as a set of line graphs. Along the horizontal axis, multiple plots depict the trend in precision, recall, normalized discounted cumulative gain, mean reciprocal rank, and hit rate as K is increased to take the model's top K predictions for the ML-100K dataset. On the top row are the metrics when performing 50x50 patch sampling, with multiple lines in each plot depicting the metrics when specifying a varying minimum density of 0.5, 0.6, 0.7, or 0.8 for the sampled patches. The bottom row of plots for each metric shows the trend in performance as K is increased to take the model's top K predictions, and compares 50x50 patch sampling with a density of 0.7 for the edges, against the performance for full graph subsampling via the novel tiling method.}
    \label{fig:edgerec_metrics}
\end{figure}

Figures~\ref{fig:edgerec_random_subgraph_metrics}~and~\ref{fig:edgerec_metrics} depict plots for the chosen suite of evaluation metrics for various top-K values. In particular, note that EDGE-Rec surpasses DiffRec baselines (included in \autoref{appendix:diffrec_metrics}) for most patch sampling tasks on edge densities above around $55\%$ (in terms of known interactions). Additionally, we observe that scaling up denoising sampling using our novel tiling method does not significantly degrade recommendations, with sufficient subgraph density.

\subsection{Model and patch size}

\begin{figure}[H]
    \centering
    \includegraphics[width=1\linewidth]{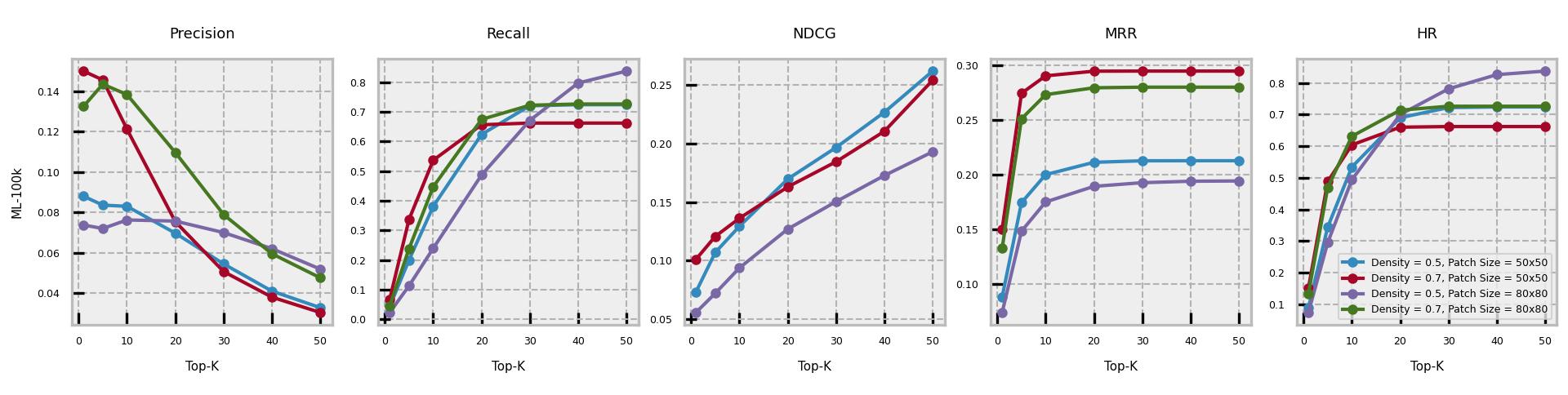}
    \caption{EDGE-Rec 8-block model evaluation metrics on ML-100k dataset. The figure depicts results on arbitrary patches of $50\times50$ and $80\times80$ sizes and specified densities. }
    \Description{Comparison of the evaluation metrics for a 8-blocks model following the previously described architecture across different choices of density and patch size plotted as a set of line graphs. Along the horizontal axis, multiple plots depict the trend in precision, recall, normalized discounted cumulative gain, mean reciprocal rank, and hit rate as K is increased to take the model's top K predictions for the ML-100K dataset. Multiple lines in each plot depict the metrics when specifying a varying minimum density of 0.5 or 0.7 and a patch size of 50x50 or 80x80.}
    \label{fig:edgerec_large_metrics}
\end{figure}

We upscale the experiments with the proposed architecture by stacking 8 GDiT blocks. \autoref{fig:edgerec_large_metrics} shows the results obtained across different choices of density and patch size. Notably, when compared with corresponding smaller baselines composed of a single block (\autoref{fig:edgerec_metrics}), we observe improvements of up to 0.13 in magnitude across all top-1, top-5 and top-10 metrics. This suggest a strong improvement in the quality of the resulting rankings overall. We further observe the role of the patch size in balancing precision and recall. Specifically, from top-20 to top-50, as the patch size increases, meaning that the number of ranked candidates is increased, \autoref{fig:edgerec_large_metrics} shows an increase in precision and recall, and a decrease in NDCG.

\subsection{Larger Datasets}

We train and evaluate a large model with 16 GDiT layers and $\sim 14$-million parameters on the larger ML-1M dataset \cite{movielens}. This dataset contains 1,000,000 ratings from over 6000 users and 4000 movies (10,000 nodes and 1,000,000 edges). A patch size of $64\times 64$ was used. Evaluation is performed by randomly sampling 10 patches of the same size from the full weighted interaction matrix, performing in-painting as previously mentioned, and averaging the metrics over these patches. Note the model predictions are compared against test edges (user ratings the model has not previously seen). 

\begin{figure}[H]
    \centering
    \includegraphics[width=1\linewidth]{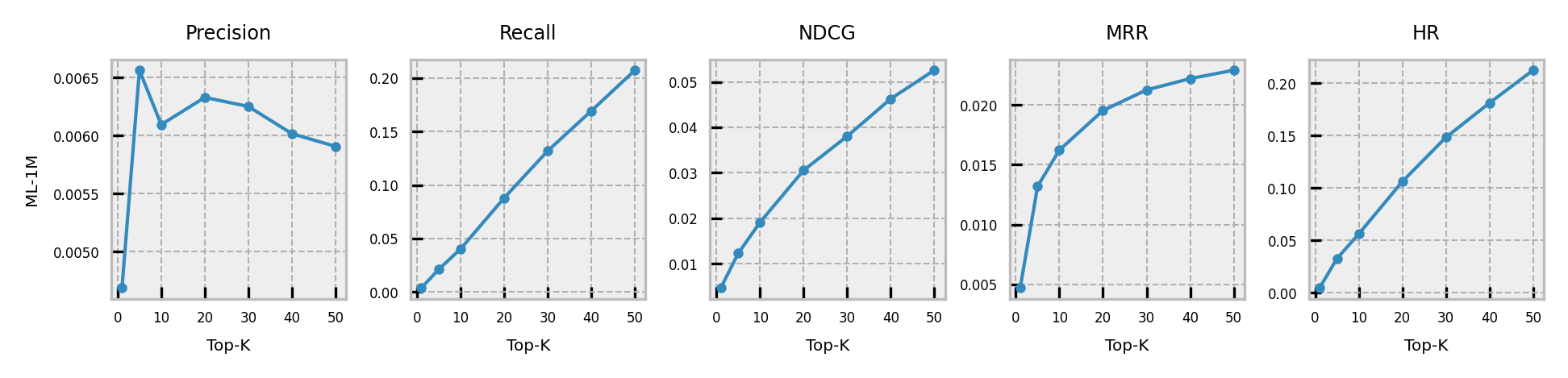}
    \caption{EDGE-Rec 16-block model evaluation metrics on ML-1M dataset. The figures show average results on arbitrary randomly sampled patches of size $64\times64$.}
    \Description{Comparison of the evaluation metrics for a 16-block model following the previously described architecture plotted as a set of line graphs. Along the horizontal axis, multiple plots depict the trend in precision, recall, normalized discounted cumulative gain, mean reciprocal rank, and hit rate as K is increased to take the model's top K predictions for the ML-1M dataset. There is one line per plot showing the trend in results for each metric as K is increased.}
    \label{fig:ml1m-results}
\end{figure}

Due to the consistency of the metrics between our evaluations on ML-100K and ML-1M, we conclude that our model is able to scale to large datasets well. Despite an order of magnitude more data, performance is relatively minimally degraded. This is most likely driven by our training strategy involving random patch sampling, while our evaluation strategy adopts this scalable approach as well.

\section{Conclusions and Limitations}

We propose a graph-diffusion-based recommendation model that predicts continuous interaction strengths. Building on advancements in conditioned diffusion models, we generate recommendations by conditioning on user and item features. By tuning the size of the patch, our model shows promising performance, despite the patch size being two orders of magnitude smaller than the complete graph. Assuming one can identify dense patches of the graph with high accuracy, this makes the training process highly scalable in the number of users and items. However, in highly sparse production-level environments, identifying such patches can be a challenge, as it essentially reduces to the retrieval problem. This makes the assumption of a dense patch size particularly strong and a clear limitation of the method. If available, future work could pair this ranking methodology with a trained retrieval model that produces high-density patches of candidate user-item pairs.

\begin{acks}
To Michael Mu, for the late-night company in the MSML lounge, and for answering our many questions on recent diffusion model developments.
\end{acks}

\newpage
\bibliographystyle{ACM-Reference-Format}
\bibliography{main}

\appendix

\newpage

\section{DiffRec Baseline Metrics}
\label{appendix:diffrec_metrics}
\begin{figure}[H]
    \centering
    \includegraphics[width=1\linewidth]{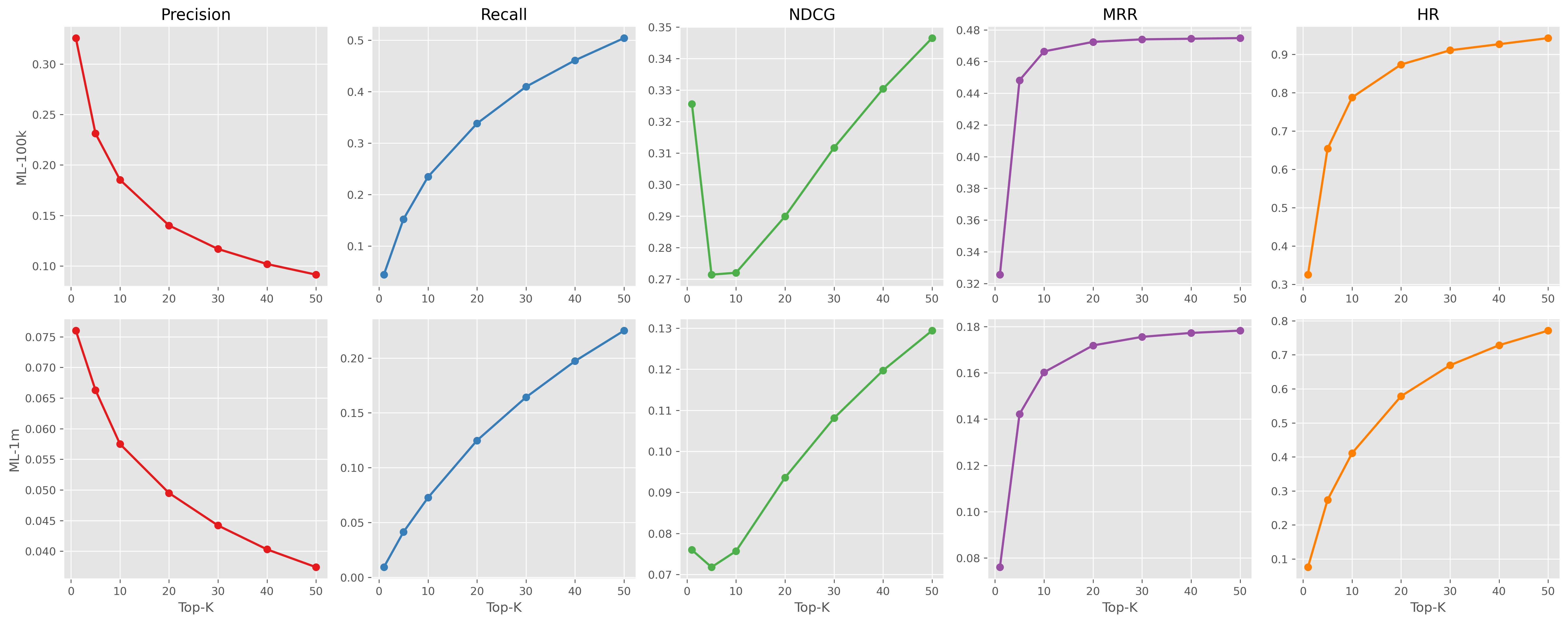}
    \caption{DiffRec \cite{wang2023diffusion} evaluation metrics on ML-100k and ML-1M.}
    \Description{Baseline metrics for the DiffRec \cite{wang2023diffusion} algorithm.}
    \label{fig:diffrec_metrics}
\end{figure}

\newpage

\section{Visualizing the Sparsity in the MovieLens datasets}

\begin{figure}[H]
    \centering
    \includegraphics[width=\linewidth]{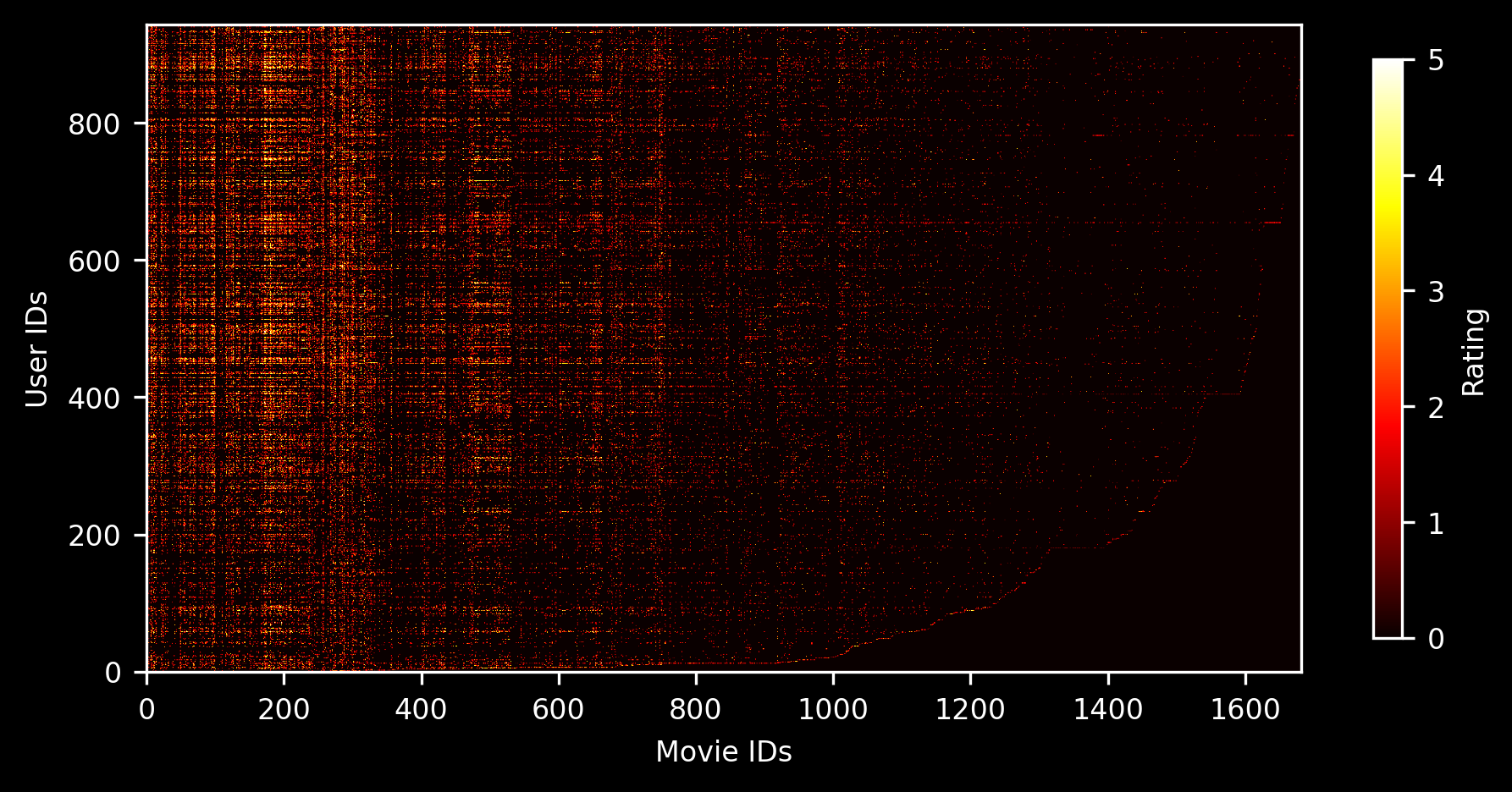}
    \caption{Heat-map of ratings provided by users for movies ranging from 1 to 5 in the MovieLens-100K dataset. Values are 0 (black) where the user has not rated the movie.}
    \Description{A plot of a heat-map for the ratings provided by users for movies in the Movie Lens 100K dataset, with an orange color intensity scheme. User IDs are on the y-axis and Movie IDs are on the x-axis. Where there are unrated movies from users, the pixels are black, and pixels are colored from red, to orange, to white corresponding to ratings of 1, 3, and 5 respectively. There is a grid like structure to the pixel colors, emphasizing popular movies that have been rated by many users, and users who have rated many movies.}
    \label{fig:movielens_heatmap}
\end{figure}

\autoref{fig:movielens_heatmap} visualizes the ratings provided by all users for the movies in the MovieLens-100K dataset. Note the inherent sparsity in the encoded interaction matrix.

\begin{figure}[H]
    \centering
    \includegraphics[width=\linewidth]{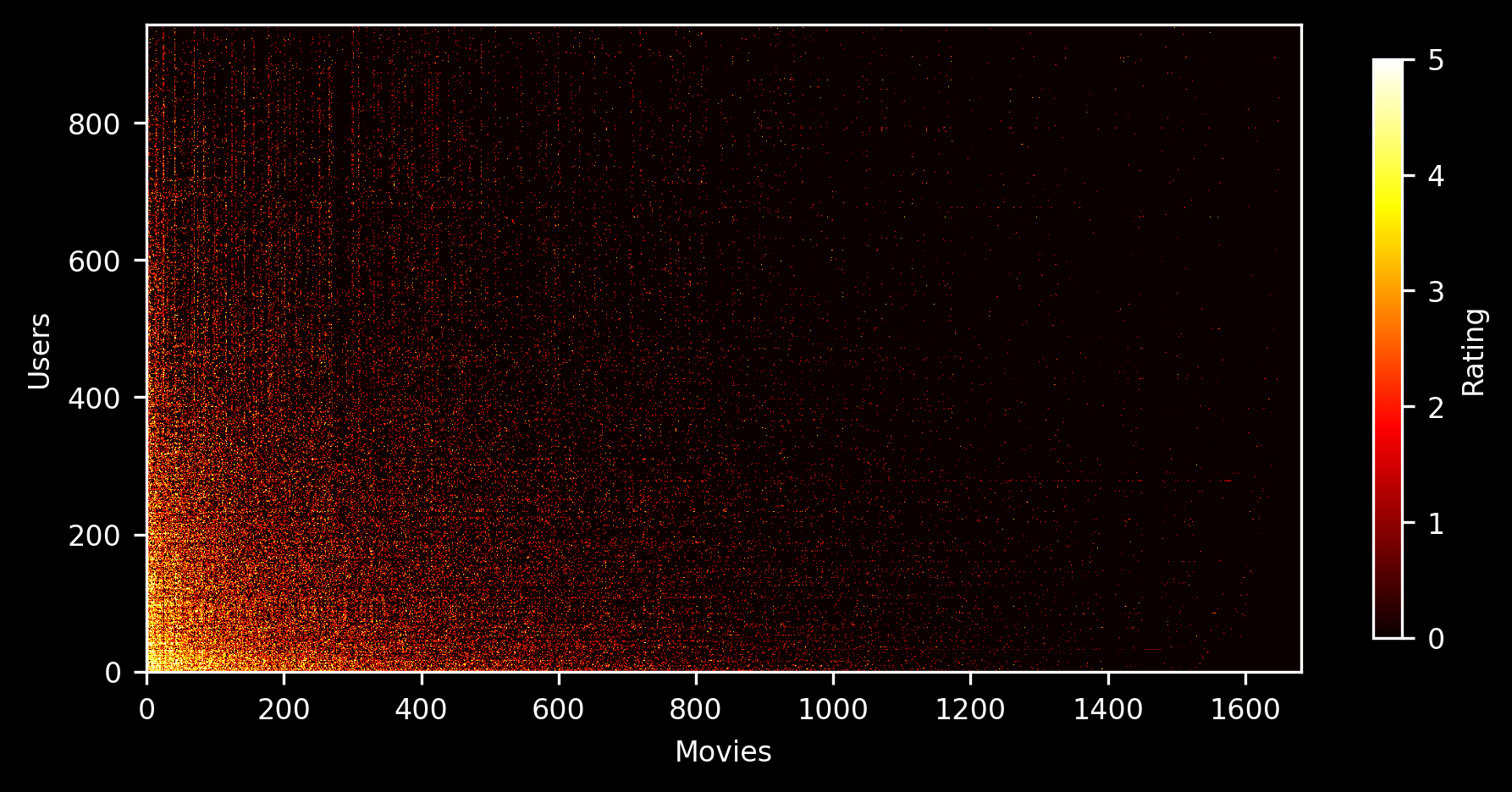}
    \caption{Heat-map of user-provided ratings for movies in the MovieLens-100K dataset, sorted heuristically by density of graph.}
    \Description{A plot of a heat-map for the ratings provided by users for movies in the Movie Lens 100K dataset, with an orange color intensity scheme. User IDs are on the y-axis and Movie IDs are on the x-axis. Where there are unrated movies from users, the pixels are black, and pixels are colored from red, to orange, to white corresponding to ratings of 1, 3, and 5 respectively. Due to the heuristic sorting based on the density of the graph, most colored pixels are concentrated near the bottom left of the plot, and the gradient going towards the bottom left changes the pixel colors from black, to red, to orange, to white.}
    \label{fig:movielens_sorted}
\end{figure}

In order to sample dense subgraphs of the interaction matrix, we heuristically identify ``important'' or densely connected users and movies by counting the number of nodes at a distance of 2 edges from each user or movie node in the interaction graph. \autoref{fig:movielens_sorted} is a heat-map of the data after sorting users and movies by this heuristic. We sample dense subgraphs of the ML-100k dataset by selecting a suitably-sized sub-matrix from the lower left corner of the visualized sorted complete graph adjacency matrix.

\end{document}